# Title: Now is the time to build a national data ecosystem for materials science and chemistry research data

**Authors:**

E.M. Campo[1,2]*, S. Shankar[3,4], A.S. Szalay[5], R.J. Hanisch[6]†

**Affiliations:**

[1]Campostella Research and Consulting, LLC.; Alexandria, VA, USA.

[2]Department of Materials Science & Engineering, University of Maryland at College Park; College Park, MD, USA.

*Corresponding author. Email: campostellaresearch@gmail.com and ecampo@umd.edu

[3]SLAC, SLAC National Accelerator Laboratory; Menlo Park, CA, USA and Department of Materials Science and Engineering, Stanford University, Stanford, CA, USA

[4]Department of Physics, Harvard University, Cambridge, MA, USA.

[5]Department of Computer Science and Physics & Astronomy, Johns Hopkins University; Baltimore, MD, USA.

[6]National Institute of Standards and Technology; Gaithersburg, MD, USA.

Campo: ORCID: 0000-0002-9808-4112

Sadas: ORCID: 0000-0002-1555-639X

Szalay: ORCID 0000-0002-4108-3282

Hanisch: ORCID 0000-0002-6853-4602

†These opinions, recommendations, findings, and conclusions do not necessarily reflect the views or policies of NIST or the United States Government

**One-Sentence Summary:**

A call for coordinated action from government, academia, and industry.

**Main Text:** It has long been argued that the economic development resulting from the integration of chemistry and materials science data in advanced manufacturing will propel the US economy through gains ranging from $123 billion to $270 billion dollars per year *(1)*. Federal funding through the last decade has produced world-class research centers throughout the country (e.g., AFLOW, CHiMaD, the Materials Project, OpenKIM, PRISMS, and others). However, these facilities largely remain "silos of excellence" and are limited in cross-platform discoverability and interoperability *(2)*. In addition, the many studies regarding research data infrastructure over the past decade have not succeeded until recently to nucleate a nationwide community and call to action. We have good examples of success stories in academia, national laboratories, and industry, but we have yet to realize a true national data ecosystem for chemistry and materials data and related research outputs. It is through a data ecosystem that the community will reach agreement on best practices and that social barriers will fall. It is then that advances in research and corresponding economic benefits will be realized.

Research organizations are critically in need of directed growth towards future interoperability and federation. In this scheme, a distributed network of data centers and data providers agree to minimum metadata standards to enable interoperability through a distributed yet federated ecosystem. We cannot emphasize this enough: it is only through community engagement and elucidation of strategic advantages that agreements will ensue. It is often said that the richest and most diverse natural ecosystems arise at the boundaries between different habitat zones. Similarly, we might expect a research data ecosystem to give rise to new discoveries at the intersections of otherwise stove-piped data repositories and services.

In 2019, a meeting was held in Chicago to discuss the need for cyberinfrastructure and to empower data-driven materials research by connecting the community. In the meeting, different domain experts gave presentations that illustrated data-driven science advances in astronomy and in the design of materials in industry. As a result, several white papers were submitted to the NSF Request for Information for several critical needs of data *(3)*. In addition, a bottom-up coalition of chemistry and materials science researchers was conceived and initiated by several of us with the purpose of improving the dissemination and re-use of materials and chemistry-related research data. The Materials Research Data Alliance (MaRDA) is a platform embracing researchers (both those that are data-savvy and those that are not) with the purpose of advancing the fields of chemistry and materials science through coordinated data management *(4,5)*.. More recently, we have observed an encouraging level of independent self-assembly and commitment and these entities would benefit from encouragement at this early stage.

The purpose of this communication is to alert government, academia, professional societies, foundations, and industry of further need for consideration of data in chemistry and materials as a long term and sustained development in the US.

**Government, Academia, and Industry**

Research in materials science and chemistry has strong stakeholders in government, academia, and private industry. Ideas that originate in academia or in-government research laboratories often convey to the private sector where they are refined and productized. However, this transition does not function very efficiently owing to the lack of common standards for research data, particularly, in the *FAIR*ness of research outputs *(6)*. In addition, the data themselves may have proprietary restrictions in industry given that the data analysis is a critical component of many industrial processes and products *(7)*. A recent report of the European Union, for example, estimated that the loss of research productivity owing to data not being *FAIR* was roughly €10B per year (*8*). The US will be challenged to compete in the international marketplace for advanced materials and chemistry unless a transition to a *FAIR*er data ecosystem is achieved.

*Cyberinfrastructure is infrastructure.*

Regardless of the workplace, access to electricity and communication technologies are taken as a given, as essential infrastructure. Access to data and associated analysis software—cyberinfrastructure—is critical to efficiency and productivity in an information-driven economy. Given the advent of advanced data processing techniques such as machine learning, it is important that the veracity and the context of the data are available for the community.

Most of the scientific output of today's large scientific instruments, built at the expense of tens to hundreds of millions of dollars, are petabytes of digital data (Large Hadron Collider, Laser Interferometer Gravitational-Wave Observatory, Spallation Neutron Source, Linac Coherent Light Source etc.). Given the scope of these experiments they will not be repeated in the foreseeable future, so their unique, high-value data sets need to be maintained and sustained, often for decades. While we understand the need to preserve physical specimens, often for centuries (as in the Smithsonian Institution and natural history museums worldwide), no such concentrated effort is under way for digital data.

Furthermore, the US cannot afford a new data center for every subdomain of science and engineering – we have to build a sharable, common data ecosystem. This is not a technological challenge, but primarily a social one: how to break down barriers and build bridges among diverse science communities, using data science as the glue.

It is frequently said that "data is the new oil" *(9)*. However, whereas oil is a limited natural resource, data are being produced at exponentially increasing rates from both experiment and simulation. This data can only lead to value in the marketplace if it can be discovered, is sufficiently documented, and is available in standard and preferably open formats. While this is currently a rare situation, it can be addressed by taking a more disciplined–and in the long run less costly–approach to data generation, management, and dissemination. Were pre-competitive data in industry more widely open and standardized we would likely see more rapid and less expensive development of new products (witness the success of the Allotrope Foundation and Pistoia Alliance in pharmaceutical companies).

*Accessible does not mean open. Open does not mean free.*

We note that making data *FAIR* does not necessarily entail that they should be either open or free; *FAIR* data are as open as possible, as closed as necessary *(10)* (In the pre-competitive world openness is highly desirable, but it is understandable that private industry will, once they have invested in productizing a fundamental research outcome, want to protect their processes and formulas. Within corporate borders, though, *FAIR* data provide an immense advantage in optimizing product development. Similarly, strategic data assets might need to be protected from sharing with nations whose policies conflict with ours in the US. The concept of platform, different from a database, is to balance the needs for proprietary use with a process of wider availability of data for the whole community.

*The role of government in the research data ecosystem.*

Federal funding, particularly from the National Science Foundation and the Department of Energy, plays a major role in materials science and chemistry research. Government agencies have the opportunity to capitalize on their investments by opening innovative programs and exploring more cost-effective approaches to grant competition, especially amongst individual PIs. We frequently see grants programs oversubscribed by factors of ten, meaning that researchers are bearing a tremendous burden in preparing proposals that are highly unlikely to be successful. As Congress considers major changes in the funding landscape across the Federal Government, grass-roots organizations such as the US Research Data Alliance (US-RDA) and the Materials Research Data Alliance can provide sound, scientific, fact-based recommendations to promote data re-use and thereby optimize the return on investment. We estimate (based upon successful examples of the recent past in astronomy and at NASA science centers) that if ~2% of research budgets were set aside for shared, open, domain-based data repositories and development of interoperability standards the challenges of building a research data ecosystem would largely be solved. Moreover, this long term vision means that information inherent in data is available across generations of scientists and engineers similar to published papers.

**Why does the US need a better strategy and appreciation for data sharing and re-use?**

Many industries in the US are capitalizing on the value and power of data from their internal processes. However, the wider research community as a whole is only slowly exploiting the capabilities of data as a critical enabler. Materials science and chemistry in the US could act as a leading demonstration of value for a national initiative, in which data are central *(11)*. It is important to note that these fields span government, academia, and importantly, the private sector, as well as a national research infrastructure that embraces these sectors in an integrated fashion that will optimize, and likely expedite, the return on investment in basic research. The Materials Genome Initiative was proposed in 2011 to accelerate the development of advanced materials through advances in computational techniques, use of standards, and enhanced data management *(12)*. This has provided key impetus to materials development both in the US and abroad *(13-16)*

There are two keys to realize high return on investment in research data: metadata and open standards. It will only be when metadata are available that discoverability and interoperability will enable data reuse at a significant scale. In addition, adoption of open data formatting standards is also essential to avoid having important metadata (instrument parameters, sample preparation processes, analysis methods) hidden behind proprietary restrictions. These keys can be secured only by building community-wide consensus.

Unfortunately, on the world stage the US is falling behind in research data infrastructure, e.g., the European Open Science Cloud (EOSC), the Global Open Science Cloud (China), and other nations that have invested in national and international capacity (Africa, Australia, France, Germany). For example, in Europe there are several activities that have been initiated and sustained for bridging fundamental analysis with the availability of data *(2, 13,15)*. At a time when the US Congress is looking into restructuring the research landscape and manufacturing is being repatriated back to the country, time is of the essence in considering a concerted national effort.

*Optimize return on investment and increase efficiency through data reuse.*

Prior funding efforts—in the hundred-million-dollar range—in materials research and chemistry have only marginally translated into a research cyberinfrastructure. Today, the US Federal Government funds chemistry and materials research in excess of $20B a year. However, most of the output data and related research products (software, simulations) are largely not curated or accessible. The situation is particularly dire in large research centers and facilities, whose program budgets are in the $10 million range per annum. These institutions lack sustainable data management plans and cyberinfrastructure, and often lack even basic data backups and access.

Lack of data—curated, documented, and *FAIR*—hinders reuse, limiting the long-term value of the initial investment. In biology, astronomy, and some other fields, success has been achieved through adoption of common standards, support for long-term repositories, and coupling of computation with data access. Examples in the biological fields include NCBI for data, PubMed for articles, etc.

But scanning the research data landscape more broadly indicates that much more needs to be done: agreements need to be reached, interfaces need to be designed and deployed. Importantly, researchers across all sectors need to appreciate the value of extant data and contribute their own datasets to the ecosystem.

Technology is changing rapidly, often faster than our big science projects. Some experiments take a decade or more, hence the data they produce will migrate to new platforms several times, but the users need to stay focused on attaining their research goals. The cyberinfrastructure we need to build has to protect end users from most aspects of these technological changes while introducing new features and paradigms that enhance their projects, becoming a "trusted intermediary" like the refereed scientific journals of today.

*Metadata are the key to discoverability and interoperability.*

While challenges in chemistry and materials research are seemingly daunting given the heterogeneity of data types, formats, experiments, instrumentation, and samples, this is indeed a feasible task. Fundamental lessons learned in other disciplines can translate into chemistry and materials research, as well as other disciplines, following established best practices for metadata development, data discovery and access protocols, and transparency in data processing and analysis tools. For example, genetics has in the past 20 years achieved discoverability and interoperability across diverse data sources. Key factors to enable data reuse include full data sets with provenance descriptors and statistically qualified data. These factors imply mature practices in the realm of calibration and automated metadata acquisition. We now know that an effective way to manage extreme data heterogeneity is to break down domains into manageable subdomains, exposing as necessary metadata that enables cross-domain functionality and long-term applicability. Efforts are coming together to introduce formats and interfaces for efficient data transfer including metadata *(16)*.

*Room for all: connecting with the world.*

Open access to research data is a vehicle for *democratizing research*. Realization of a cyberinfrastructure in this realm opens opportunities to those who are otherwise unable to contribute given the high cost of experimental apparatus and high-performance computation. Minority serving institutions and community colleges in the US often fall in this category. A nation-wide cyberinfrastructure will enable data sharing, use, and discovery by all participants.

There is room in this research environment for international data organizations, such as the "Data Together" four: Research Data Alliance, CODATA, World Data System, and GO-*FAIR*. There is also room in this research environment for scholarly societies and their publishing partners, for public and private funders, and for research libraries, as well as domain repositories, generic repositories, and service providers such as DataCite and ORCID. All are welcome at the table. In fact, they all are needed at this table in order to realize the goal of a *FAIR* research ecosystem in the US that is capable of safely connecting to similar and synergistic efforts globally. Of course, being *FAIR* does not require being open; we need to protect national interests be they economic or security related. But on whole, openness leads to better and more robust research, allowing full scrutiny and truly evidence-based advances in science and science policy.

**Beyond the Vision: Immediate Actions Needed**

A number of comprehensive studies have been conducted and we now have an opportunity to act based on the recommendations resulting from their analysis (17). A decade's worth of these intelligence reports identifies a federated infrastructure as a requirement but in which only minor progress has been made. What specifically is missing and how can we go about rectifying the situation?

What is needed is a distributed but federated network. This network builds upon data management principles set forth by organizations such as the Global Research Data Alliance, CODATA, GO-FAIR, and the World Data System. Data will be distributed in order that they are curated by domain experts, with appropriate funding. Once curated, data needs to be federated through agreements on both data discovery and access protocols such as those pioneered by Optimade (https://optimade.org). These international data organizations will play a crucial role in assuring adequate diversity, training, and mentoring (perhaps with the younger generations guiding the more senior!). Participants and stakeholders in the national ecosystem will be aware of their rights and responsibilities through the nascent NIST Research Data Framework (RDaF) *(17, 18)*. Indeed, the RDaF has identified materials and chemistry research community as an initial pilot study. It is particularly important to understand the role of commercially sponsored research data repositories, lest publicly funded data become locked behind paywalls. The example of a Wikipedia-type frame for the data and the meta-data maybe something to consider, where critical information is available to the entire community efficiently.

*Where do we start? Science first.*

In the information era, the scientific questions and anticipated socio-economic impact of the research funding should take priority. Today, we see topical efforts around microstructures, 2D materials, crystal growth, thermoelectrics, and quantum materials as critical areas for cyberinfrastructure investment towards maintaining US industrial competitiveness. Scientific progress will derive from capacity building through metadata standards, *FAIR* communities of practice, and incentives that need to be nurtured throughout the ecosystem. We are confident that community building will generate consensus-based, grass-roots practices that will be self-enforced throughout the ecosystem by way of awards versus rejections, hence avoiding the need of additional policing.

*Now is the time.*

It has been increasingly clear that uncoordinated funding efforts throughout the federal landscape are not conducive to efficient and results-oriented outcomes. This could not be any clearer for research data capacity building; a sound, cohesive, and highly performing infrastructure requires concerted strategic funding from all government agencies We have an opportunity to set US data policy and practice on a new and improved path, building on the Federal Data Strategy and the Evidence Act but expanding scope beyond unfunded mandates into the construction of a truly *FAIR* and democratic data and computational ecosystem that benefits government, academia, and industry equally. We can validate this approach in the materials science and chemistry domains, then expand to embrace US scientific research more broadly under the rubric of an Open Research Commons. If we do not accept this challenge, the lost opportunity cost will be immense, estimated at €10B/year in Europe alone *(8)*. And we have critical research challenges ahead of us, e.g., in quantum information science, where we will as a nation fall behind without data sharing infrastructure.

In conclusion, action is needed on several fronts:

Mobilizing government agencies to coordinate data and computation initiatives.
- NSF, NASA, DOE, DOD, NOAO infrastructure is all separate and independent
- HPC systems and cloud platforms are not interoperable
- Materials science and chemistry could be leadership communities within a US Open Research Commons
- NIST Research Data Framework is needed to describe the landscape, roles, and responsibilities
- Establish a US leadership office to coordinate
- Implement the Federal Data Strategy and Evidence Act

Dedicating resources to research data infrastructure.
- ~2% of research investment
- Development of metadata standards
- Wide use of open data formats
- Common discovery and access protocols

Improving incentives.
- Recognition for data sharing
- Expanding criteria for promotion and tenure
- Improving the return on investment in basic and applied research

We note that an open research data ecosystem enables equity:
- Democratizes access
- Engages underserved communities through better access to data and computation

Success requires participation from many stakeholders:
- Funding agencies
- Philanthropies
- International data organizations
- National and international research federations
- National Academies
- Scholarly publishers
- Professional societies
- Academia
- Industry

Now is indeed the time to establish a national strategy and concomitant infrastructure focused on research data.  What are we waiting for?

**References and Notes**

1. T. Scott et al. 2018, https://www.nist.gov/system/files/documents/2020/02/06/MGI%20Final%20Report.pdf.
2. L. Himanen et al. 2019, https://onlinelibrary.wiley.com/doi/pdf/10.1002/advs.201900808
3. https://www.nsf.gov/cise/oac/datacirfi/rfi_responses.jsp
4. MaRDA, https://www.marda-alliance.org/.
5. RDA-US, https://us.rd-alliance.org/.
6. Wilkinson, M., Dumontier, M., Aalbersberg, I. *et al.* The FAIR Guiding Principles for scientific data management and stewardship. *Sci Data* 3, 160018 (2016). https://doi.org/10.1038/sdata.2016.18.
7. Materials Scientists Look to a Data-Intensive Future, Science  23 Mar 2012: Vol. 335, Issue 6075, pp. 1434-1435.
8. http://publications.europa.eu/resource/cellar/d375368c-1a0a-11e9-8d04-01aa75ed71a1.0001.01/DOC_1
9. https://www.forbes.com/sites/forbestechcouncil/2019/11/15/data-is-the-new-oil-and-thats-a-good-thing/.
10. https://www.go-fair.org/resources/faq/ask-question-difference-fair-data-open-data/.
11. The Fourth Paradigm: Data-Intensive Scientific Discovery, T. Hey S. Tansley K. Tolle, Microsoft Research, October 2009.
12. Materials Genome Initiative for Global Competitiveness; National Science and Technology Council, Office of Science and Technology Policy: Washington, DC, 2011
13. Ghiringhelli, L. M. et al. Towards efficient data exchange and sharing for big-data driven materials science: metadata and data formats. npj Comput. Mater. 3, 46 (2017).

14. de Pablo, et al., New Frontiers for the Materials Genome Initiative. npj Computational Materials 2019, 5, 41.

15. Huber, S.P., Zoupanos, S., Uhrin, M. *et al.* AiiDA 1.0, a scalable computational infrastructure for automated reproducible workflows and data provenance. *Sci Data* 7, 300 (2020). https://doi.org/10.1038/s41597-020-00638-4

16. J. M. Cole, "A design-to-device pipeline for data-driven materials discovery," Acc. Chem. Res. 53, 599–610 (2020).

17. https://doi.org/10.6028/NIST.IR.8364

18. RDaF, https://www.nist.gov/programs-projects/research-data-framework-rdaf.

**Author contributions:**

Conceptualization: EMC, RJH

Writing – original draft: EMC, RJH

Writing – review & editing: EMC, RJH, SS, AS

**Competing interests:** EMC is a member of the Research Data Alliance and has recently worked to establish the RDA Region of the Americas in scientific, ambassadorial, and strategic capacities. RJH is an elected member of the RDA Council.